\documentclass{mpi}

\title{On the effectiveness of virtualization-based security}
\author{}

\usepackage{graphicx}

\newcommand{\helloRootkitty}{\emph{Hello Rootkitty}}
\newcommand{\HelloRootkitty}{\emph{Hello Rootkitty}}

\begin{document}

\maketitle
\begin{abstract}



Protecting commodity operating systems and applications against malware and targeted attacks has proven to be difficult. In recent years, virtualization has received attention from 
security researchers who utilize it to harden existing systems and provide strong security guarantees. This has lead to interesting use cases such as cloud computing where possibly 
sensitive data is processed on remote, third party systems. The migration and processing of data in remote servers, poses new technical and legal questions, such as which security 
measures should be taken to protect this data or how can it be proven that execution of code wasn't tampered with.
In this paper we focus on technological aspects. We discuss the various possibilities of security within the virtualization layer and we use as a case study \HelloRootkitty{}, a 
lightweight invariance-enforcing framework which allows an operating system to recover from kernel-level attacks. In addition to \HelloRootkitty{}, we also explore the use of special 
hardware chips as a way of further protecting and guaranteeing the integrity of a virtualized system.


\end{abstract}


\section{Introduction}
Virtualization is the set of technologies that together allow for the existence of multiple running virtual machines on-top of a single physical machine. While initially all of the needed mechanisms for virtualization were created in software, the sustained popularity of virtualization, lead to their implementation in hardware, providing the desired speed that was lacking in their software counterparts. Today virtualization is gaining more and more interest in IT as well as the business world. The ability of virtualization to easily consolidate and migrate virtual machines between physical machines, allows corporations to outsource their IT infrastructure while reducing the cost of maintenance, power consumption and required infrastructure. This practice is called {\em cloud computing}. While virtualization is now mostly a technology
for server farms, it is expected that it will soon affect both the mobile device and desktops market. Companies are already offering products such as VMWare's Mobile Virtualization platform~\citep{Barr:2010} and Citrix's XenDesktop which utilize virtualization to increase productivity, decrease costs and allow easier maintenance of mobile and desktop devices in corporate environments.

From a security point of view, researchers have already identified a number of attractive properties of virtualization that can be used to provide stronger security guarantees in
server, desktop as well as mobile environments. The property that has received the most attention is the guaranteed isolation between virtual machines on top of the same physical machine and
the isolation of the code managing the virtual machines, called the hypervisor, from the virtualized operating systems. This makes the hypervisor and separate virtual machines ideal locations for security measures. A strong isolation between the layer where a security mechanism resides and the layer which it protects allows the security mechanism to continue to operate correctly even in the presence of an attack against the protected layer.
A number of virtualization-utilizing security mechanisms have been proposed, ranging all the way from rootkit detectors to anti-virus products~\citep{Qubesos, Gadaleta2009, hellorootkitty,chiueh:sade}.
Unfortunately virtualization-utilizing security measures are usually affected by consistent overhead since they typically require interactions between the hypervisor and the virtualized operating
system that would not be present in traditional environments.


In this paper we discuss two very active research tracks. First, one track focuses on the application of virtualization techniques to increase the security of the overall system against malicious software. As a use case we describe \helloRootkitty{}, an in-hypervisor invariance-enforcing framework used to detect kernel-level rootkits. While \helloRootkitty{} significantly elevates security of the overall system without requiring any hardware or software modification, it is not able to guarantee isolation of sensitive information against a determined attacker.

Second, we discuss recent research results of another research track to provide formally provable security for small, specially tailored software modules. These strong security guarantees come at a significant cost of partitioning applications in security sensitive and insensitive parts. These results build on the technology of the Trusted Platform Module (TPM), a low-budget chip that is currently shipping with newer computers that provides a limited number of security features in hardware.

The rest of the paper is structured as follows. Section~\ref{sec:rethinking} discusses how virtualization can be used to rethink security from the ground-up and use it
to develop strong defenses for modern computing. Section~\ref{sec:helloroot} explores \helloRootkitty{} and shows how a hypervisor can be used as an invariance enforcing
framework in order to identify malicious and unexpected modifications in a kernel's data structures. Section~\ref{sec:tpm} presents the technology of TPMs and discusses how this hardware chip can be employed to offer provable security, followed
by our conclusions in Section~\ref{sec:conclusion}

%

\section{Rethinking security}
\label{sec:rethinking}

In order to take advantage of cloud computing in a corporate environment, a strong isolation between workloads of different parties is required. Modern operating systems, such as Microsoft Windows Server 2008, are not able to provide sufficient isolation between different applications, or even between applications and the operating system itself. According to the National Institute of Standards and Technology (NIST), that keeps track of all reported vulnerabilities of commercial available software packets, there were 128 new vulnerabilities found in Windows Server 2008 in 2010 alone \citep{nist2011}. For 2011, this figure was already surpassed at the time of writing. The size of the kernel, ranging into millions lines of code, make it infeasible to make it reliably secure. Subtle bugs\citep{alephOne1996smashing} can be exploited by an attacker to gain kernel-level access. As this is the most privileged level, sensitive information stored by another party's applications can now be easily accessed.

Virtualization techniques, allow us to create another, even more privileged, layer. Using this additional layer, research has focused on two distinct approaches. First, the layer can be used to offer stronger protection of the operating system running on top of it without any need to modify any source code or binary. Second, security measures can be implemented in this layer to protect the execution and isolation of code even in the presence of malware. While the latter approach is able to offer stronger security guarantees, it requires a significant modification of source code.

\subsection{Hello Rootkitty: an invariance-enforcing framework}
\label{sec:helloroot}
Rootkits are pieces of malicious software deployed on a compromised operating system with the chief purpose of concealing the presence of other malicious applications (such as a keylogger, or 
a backdoor) from the users and administrators of that system. The two most common rootkit classes are user-mode and kernel-mode and essentially signify the privilege level where the rootkit resides.
User-mode rootkits have a relatively limited impact on the system because they compromise a single application at a time and can be easily
detected and removed by security mechanisms residing either in the userspace or the kernel-space of the operating system. Kernel-mode rootkits however, are much more insidious, with a higher 
impact on the system and harder to detect and remove. 

A countermeasure deployed within the same layer of the system that it protects might be circumvented and is susceptible to attacks. No isolation can be guaranteed if the countermeasure that protects
and the kernel that is to be protected, are both part of the attack surface. \emph{Hello Rootkitty} takes advantage of isolation provided by virtualization in order to protect a target kernel and mitigate the problem of rootkits.
We assume that a rootkit can be introduced in a system through a Loadable Kernel Module (LKM) \footnote{LKMs are regularly used to install new hardware or extend the kernel with new features. Unfortunately, an inexperienced user can easily install a malicious LKM which masks itself as a benign application.}, by overwriting memory directly via kernel-exposed interfaces, or by exploiting a vulnerability in the kernel that allows execution of arbitrary code. 
A common characteristic of most rootkits is that they overwrite locations in memory in order to change the control-flow inside the kernel. The majority of these locations have 
values that do not change during normal execution. Thus, any sign of variance can be used to detect the presence of rootkits. We name these target memory areas as invariant ``critical kernel objects'' because compromising such locations is essential to change the control-flow of the kernel and execute injected code.

The literature provides methods to detect invariant critical kernel objects as described in \cite{6,7,8,HookSafe}. These methods differ depending on the type of kernel object. We identify three types of objects:

\begin{enumerate}
\item Static kernel objects at addresses hard-coded and not dependent on kernel compilation
\item Static kernel objects whose addresses depend on kernel compilation 
\item Dynamic kernel objects allocated on the kernel heap via kernel-specific memory allocation functions
\end{enumerate}

Once the locations of invariant kernel objects have been collected, \HelloRootkitty{} can check their integrity regardless of their type. The minimal information 
required to enable protection without dealing with false positives is the address of the object within guest memory and its size in bytes.

Part of \HelloRootkitty{} is a trusted module which operates in the guest operating system at boot time and provides such information to the hypervisor. We consider boot time 
our root of trust. This is a realistic assumption especially in the case of production servers, such as mail and web servers, in which the environment does not change after their installation. 
After the first boot, the system is considered to operate in an untrusted environment and integrity checking will be enforced by the hypervisor.

\begin{figure}[htbp] 
\begin{center}
\includegraphics[scale=0.5]{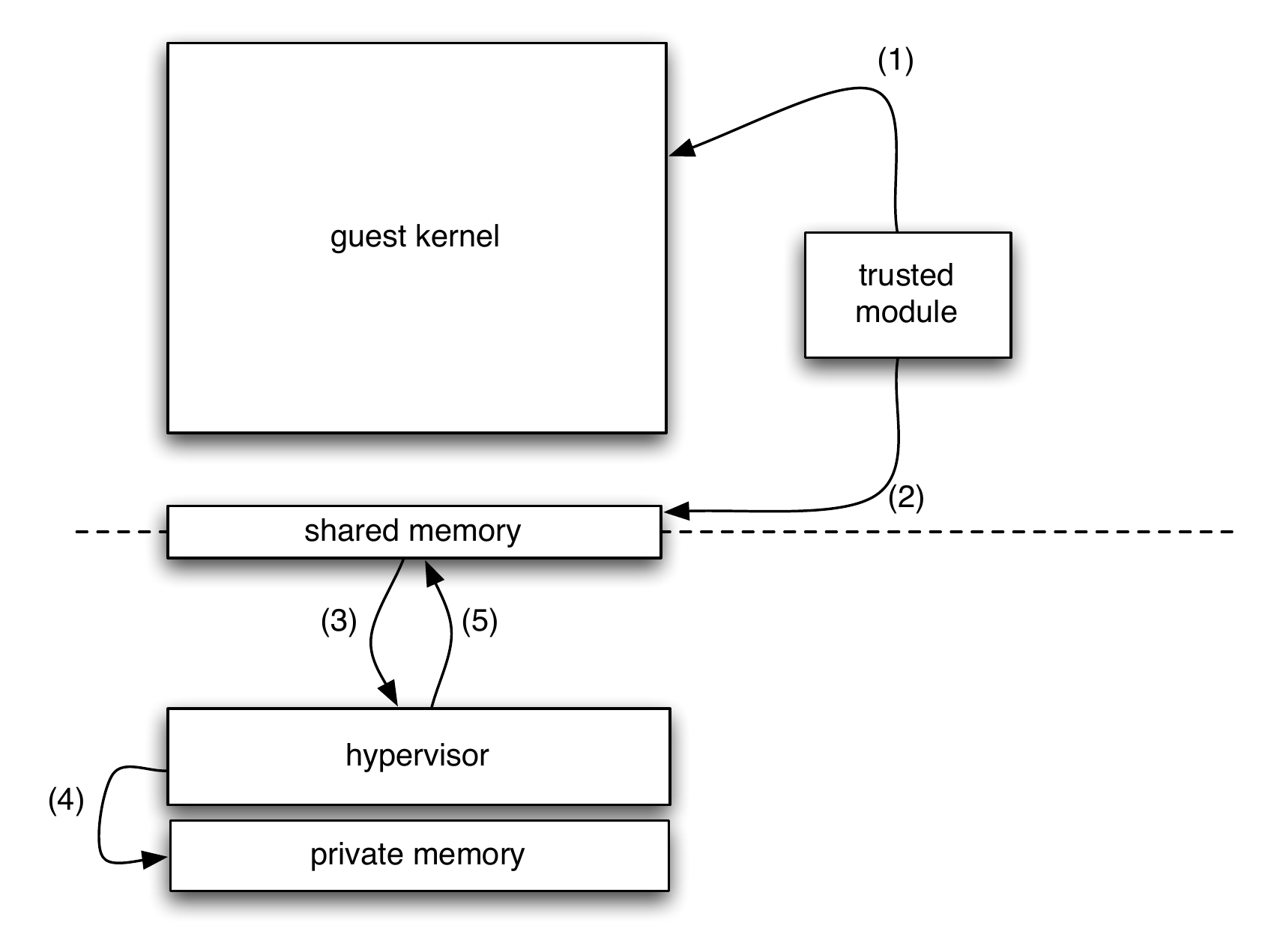}
\caption{{\bf High level view of trusted module-hypervisor interaction in \HelloRootkitty{} framework}}
\label{schema_tm_hyperv}
\end{center}
\end{figure}

A schema of \HelloRootkitty{} is provided in Figure \ref{schema_tm_hyperv}. 
Given the list of invariant kernel objects, the trusted module sends this data to the hypervisor via a hypercall\footnote{Virtual machines communicate to the underlying hypervisor via hypercalls, the equivalent of system calls used by regular processes to communicate with the underlying kernel.}. The hypervisor will checksum the contents at the provided addresses, store the computed hashes to a private memory area, not accessible by the guest, and will force the trusted module to unload. After this point, an attacker can no longer tamper with the countermeasure: the trusted module 
is not part of the attack surface and isolation between the guest and the hypervisor is guaranteed by virtualization-enabled hardware. 

Integrity checking is needed to detect if any of the protected objects has been compromised. In order to detect such changes, the hypervisor needs to access 
the contents within the guest, compute their hash and finally compare it against the hash stored in its private memory area.

\HelloRootkitty{} executes checking by taking into account the regular interaction of the hypervisor and the guest running on top. 
In a virtualized environment the guest runs on a logical processor in a privilege level lower than the virtualized machine, called \texttt{VMX non-root} mode. In this mode certain instructions or events triggered by the guest kernel will cause a 
\texttt{VMExit} and control is given to the hypervisor. The hypervisor will handle the exception and return to the guest upon termination. 
We found that writing to control registers\footnote{MOV CR* for Intel x86 Architecture} is the most convenient event to check the integrity of kernel objects. This event is strategic because with virtual addressing enabled, Control Register 3 (CR3) becomes the page directory base register. On Intel architecture switching between two running processes will change CR3. This gives direct information about the current guest system load and allows to implement a countermeasure that scales accordingly.

When the hypervisor detects that the signature of a protected object does not match the one computed the first time, the system will report an ongoing attack. \HelloRootkitty{} is also 
able to repair the compromised object, if a copy of it has been provided by the trusted module. 

Executing integrity checking outside of the target operating system can have consistent cost that affects performance overhead and limit the deployment in production systems. Moreover, the number of critical kernel objects to be protected is usually high and checking the integrity of the entire list could make the system unusable. 
Thus, integrity checking of a large list of objects is spread over a certain number of events trapped by the hypervisor. This problem relaxation considerably improves the performance overhead, although it comes at a cost in terms of security and detection time. However, the detection ability of this countermeasure remains strong also in a realistic scenario.

While performance benchmarks show negligible overhead and the memory footprint is proportional to the number of protected objects, \emph{hello Rootkitty} has some limitations. 
Since it depends on invariance inference engines to provide an accurate list of invariant critical kernel objects, \emph{hello Rootkitty} will be unable to detect attacks that occur in the non-reported ones. Moreover, this framework will not enable protection of objects whose values can legitimately change during regular usage. 
However, protecting a consistent amount of invariant kernel objects will dramatically reduce the attack surface and make the overall system more secure against rootkits.

\subsection{Offering provable secure isolation}
%

\label{sec:tpm}
While \HelloRootkitty{} significantly elevates the security of the overall system, it is not able to guarantee complete isolation of applications. Malware could still successfully exploit subtle bugs in the kernel by trying to restore invariants before they are checked or avoiding breaking invariants altogether. For at least a certain period of time, the malware may be able to access sensitive information of applications running on behalf of other parties.

An alternative research track attempts to protect sensitive information even in the presence of kernel-level malware. Security measures have been developed\citep{garfinkel2003terra,singaravelu2006ReducingTCB,mccune2008Flicker,mccune2010TrustVisor,strackx2010trustedSubsystemsInEmbeddedSystems,azab2011sice} to offer such strong security guarantees by splitting applications in a security sensitive and a security insensitive part. The former part is executed in complete isolation from the rest of the system. This minimizes the size of the trusted computing base (TCB), the sum of all software that is relied upon to isolate sensitive information and calculate the correct result, up-to a point that it becomes feasible to formally verify\citep{jacobs2008verifast} the correctness of code. This proves with mathematical certainty that sensitive information will never leak to another party. While these security measures are able to provide very strong security guarantees, they are no longer binary-compatible with legacy applications. A significant effort is required to partition these applications in an security sensitive and insensitive part. Moreover, it does not provide any availability guarantees. While malware is unable to access or modify sensitive information, it still can, for example, cause the system to freeze, preventing all parties to execute any application.

\subsubsection{Root of trust}
%

Accepting the presence of kernel-level malware causes significant problems to establish a root of trust. Malware may already have infected the system before the security measure is applied. Hence, the malware could influence the security measure's behavior or disable it altogether. For example, the kernel image stored on the hard drive, could have been modified to include the malicious code. To mitigate this problem, a low-cost security chip, called the Trusted Platform Module (TPM)\citep{TCG2004TPMdesign} has been developed and is already shipped with most modern computers. Equipped with a slow but cheap processor and its own memory, it can be used to execute a fixed set of security-related tasks. To protect the chip itself from software attacks, it is shipped with all required software that under no circumstance can be modified.

The TPM chip is designed for a few specific tasks. First, it is able to record all software that is loaded on the system. Starting at power up, a measurement of the software is calculated and stored on the chip. Every time a new process is loaded, this measurement is extended with a measurement of the loaded software including the used configuration files. This is called a {\em Static Root of Trust Measurement} (SRTM). Similarly, a new measurement can be started after the system has already booted, called a {\em Dynamic Root of Trust Measurement} (DRTM).
Second, the TPM chip is able to store a very limited amount of data, called {\em sealed storage}. On storage, the data is supplied together with a measurement. Only when software with this specific measurement is loaded, can the data be retrieved again.
Finally, the chip is able to attest to a third party that a specific version of software has executed and outputted the specified results. Using cryptographic functions, it can prevent malware from making false claims such as specifying a different output.


%

Using a combination of the features directly provided by the TPM chip, strong security guarantees can be provided. Flicker\citep{mccune2008Flicker}, takes this approach. Applications are divided into security sensitive modules. Upon invocation, a new dynamic root of trust measurement is started, measuring the loaded module. This will also place the machine in an isolated state. During that time, only the module itself has control over the machine. Malware, possibly already present on the system, will not be executed. Hence, it is unable to access any sensitive information used by the module. After the module finished its execution, it will clear all sensitive information from memory and resume normal operation of the system. When information needs to be stored for other invocations of the same or other modules, it uses the sealed storage feature of the TPM chip.

While malware is not able to modify the binary image of the module, the module must still be trusted. Subtle bugs in the module itself, may be exploited by an attacker by providing unexpected parameters. As a result, sensitive information may not be completely overwritten after the module finished its execution or an incorrect result may be provided. Hence, there is still a need to formally verify that modules behave correctly \citep{jacobs2008verifast}. Given the very limited code size, contrary to entire monolithic kernels such as Windows and Linux, this approach is feasible.

While offering strong security guarantees, the Flicker security measure still has significant drawbacks. First, isolating security-sensitive parts of an application in modules can be difficult. {\em All} accesses to sensitive information must be encapsulated in the module in such a way that the result does not reveal any information unknown to an attacker. This could be achieved by, for example, encrypting the provided result. Another difficulty is that security-sensitive functionality provided by the operating system can no longer be used by the module. 
Second, to reduce the cost of the TPM chip, it is equipped with a low-budget processor that is much slower than the main processor of the system. As a result, executing a module incurs a significant overhead. Moreover, to isolate the module, other code is prevented from being executed and the user can experience a momentarily freeze of the system.

\subsubsection{Increasing flexibility}

Recent security measures\citep{mccune2010TrustVisor,sahita2009criticalAppsOnMobile} are able to significantly reduce Flicker's drawbacks while offering the same strong security guarantees. Using virtualization techniques, an additional protection layer can be added. This hypervisor layer is even more privileged than kernel layer and can be used to protect against kernel-level malware. As usual, formal verification is required to avoid malware infecting this most privileged level.

TrustVisor is an example of a security architecture that uses virtualization techniques to offer strong isolation of code and data of sensitive parts of an application. When the system is booted, a hypervisor is installed. This code executing at the highest privilege level, has two purposes. First, it is responsible to maintain binary compatibility with legacy code. Given the number and diversity of operating systems and legacy applications, any security measure that requires even minor changes to legacy software are infeasible in practice. Only applications that require use of the newly offered security guarantees should require minimal modification.

Second, the hypervisor must protect memory regions used by itself or protected modules. To be able to guarantee isolation of modules, TrustVisor must prevent read and write access from malware to these memory regions from the legacy operating system or applications. Similarly, protected modules that are possibly specially crafted by an attacker must not be able access other protected modules or the implementation of the security measure itself.

Using these properties, TrustVisor implements and protects software-based TPM's, called $\mu$TPMs, to significantly increase performance. When the system is booted, the hardware TPM chip measures all loaded software. At the first execution of the security measure, a long term secret is created for the $\mu$TPMs and sealed by the TPM. For subsequent boots, only when the security measure is loaded correctly, access to the long-term secrets is granted. The $\mu$TPMs use this long-term secret to safely provide secure storage and attestation functionality without accessing the slow hardware TPM. When a security-sensitive part of an application request TPM functionality, a $\mu$TPM is accessed instead. This reduces the overhead of accessing the hardware TPM chip to the cost of crossing the kernel-hypervisor border. Since $\mu$TPMs are executed on the main processor which is significantly more performant than the low-cost hardware TPM, performance evaluation of TrustVisor shows a significant speedup compared to Flicker.


\section{Conclusion}
\label{sec:conclusion}
%

Recent years corporations have been looking at cloud computing as a convenient model to outsource their IT infrastructure. The key technology to enable this is virtualization. It allows the execution of several virtualized machines on the same physical hardware decreasing the cost of maintenance, power consumption and required infrastructure. Virtualization, when supported by hardware, also comes with important features that can be used to implement strong security measures such as isolation and a low performance overhead.

In this paper we described two distinct active security research tracks. First, as a use case, we presented \helloRootkitty{}, a lightweight security measure that mitigates the problem of kernel level rootkits. Upon detection of malware \helloRootkitty{} alerts the administrator of the virtual machine and, in some cases, proceeds to repair the compromised kernel. This security measure has been implemented within a hypervisor. Using virtualization techniques, it doesn't require any modification of the target kernel nor of any legacy user application. Moreover performance overhead is low enough to be applicable to a large spectrum of systems. Although it significantly elevates security, it can't guarantee full isolation of applications and can leave sensitive information unprotected.

Secondly, we described recent results of another active research track that focuses on provable protection of sensitive information. The core idea of security measures in this research field, is to partition target applications into a security sensitive and an insensitive part. This is achieved by modifying the target application according to a detailed application specific functional analysis. For most use cases a significant effort is required. Moreover to achieve complete isolation from malware, a root of trust needs to be established by supported hardware.

Most countermeasures originating from these two research tracks are complementary. Strong isolation of sensitive information and availability guarantees can be provided, even when multiple, possibly malicious, parties are executing on the same platform, as is the case of cloud computing. 

\paragraph{Acknowledgements:}
This research is partially funded by the Interuniversity Attraction Poles Programme Belgian State, Belgian Science Policy, IBBT, the Research Fund KU~Leuven, the Flemish agency for Innovation by Science and Technology (IWT) and EU FP7 project NESSoS.

\bibliography{paper}

\end{document}